\documentclass[twocolumn]{revtex4}
\usepackage{mathbbold}
\usepackage{amsfonts}
\usepackage{amsmath}
\usepackage{amssymb}
\usepackage{charter}
\usepackage{graphicx}

\input{tcilatex}
\begin{document}

\title{Generating Hermite polynomial excited squeezed states by means of
conditional measurements on a beam splitter}
\author{Xue-xiang Xu$^{1,\dag }$, Hong-chun Yuan$^{2}$ and
Hong-yi Fan$^{3}$}
\affiliation{$^{1}$College of Physics and
Communication Electronics, Jiangxi Normal
University, Nanchang 330022, China\\
$^{2}$College of Optoelectronic Engineering, Changzhou Institute of Technology, Changzhou 213002, China\\
$^{3}$Department of Material Science and Engineer, University of Science and Technology of China, Hefei 230026, China\\
$^{\dag }$Corresponding author: xxxjxnu@gmail.com }

\begin{abstract}
A scheme for conditional generating a Hermite polynomial excited squeezed
vacuum states (HESVS) is proposed. Injecting a two-mode squeezed vacuum
state (TMSVS) into a beam splitter (BS) and counting the photons in one of
the output channels, the conditional state in the other output channel is
just a HESVS. To exhibit a number of nonclassical effects and
non-Guassianity, we mainly investigate the photon number distribution,
sub-Poissonian distribution, quadrature component distribution, and
quasi-probability distribution of the HPESVS. We find that its
nonclassicality closely relates to the control parameter of the BS, the
squeezed parameter of the TMSVS, and the photon number of conditional
measurement. These further demonstrate that performing the conditional
measurement on a BS is an effective approach to generate non-Guassian state.

\textbf{ocis: } (270.5570) Quantum detectors; (270.4180) Multiphoton
processes; (270.5290)  Photon statistics

\end{abstract}

\maketitle

\section{Introduction}

Quantum state engineering has been a subject of increasing interest to
construct various novel nonclassical states in quantum optics and quantum
information processing\cite{a1,a2}. From a theoretical point of view, the
simplest way of generating nonclassical field states is to apply the photon
creation operation to classical states such as the thermal and coherent
states\cite{a3,a4,a5}. These nonclassical states, such as the single-photon
added coherent state\cite{a6} and single-photon-added thermal state\cite{a7}%
, have been realized experimentally. Subsequently, it has been demonstrated
that subtracting photons from traditional quantum states exhibit an
abundance of nonclassical properties\cite{a8,a9,a10,a11}. Photon subtraction
or addition can improve entanglement between Guassian states\cite{a12},
loophole-free tests of Bell's inequality\cite{a13}, and quantum computing%
\cite{a14}.

To meet the requirement of the development of quantum optics and quantum
information tasks, some nonclassical states are explored by performing the
different combination of photon subtraction and photon addition\cite%
{a15,a16,a17,a18,a19,a20}, which have different properties. Kim et al\cite%
{a21} discussed single photon adding then subtracting (or single photon
subtracting then adding) coherent state (or thermal state) to probe quantum
commutation rules $\left[ a,a^{\dag}\right] =1$. Lee et al\cite{a17}
investigated the nonclassicality of field states when photon
subtraction-then-addition operation or the photon addition-then-subtraction
operation is applied to the coherent state (or thermal state), respectively.
Yang and Li\cite{a18} analyzed multiphoton addition followed by multiphoton
subtraction ($a^{l}a^{\dag k}$) and its inverse ($a^{\dag l}a^{k}$) on an
arbitrary state. Recently, Lee and Nha\cite{a23} proposed a coherent
superposition of photon addition and subtraction, $ta+ra^{\dag}$ ($%
\left
\vert t\right \vert ^{2}+\left \vert r\right \vert ^{2}=1$) acting on
a coherent state and a thermal state. More recently, we investigated the
nonclassical properties of optical fields generated by Hermite-excited
coherent state\cite{a24} and Hermite-excited squeezed thermal states\cite%
{a25}.

On the other hand, another promising method for generating highly
nonclassical states of optical fields is known to be conditional measurement%
\cite{a25a,a26,a27,a28,a28a}. Namely, when a system is prepared in an
entangled state of two subsystems and a measurement is performed on one
subsystem, then the quantum state of the other subsystem can be reduced to a
new state. In particular, it turned out that conditional measurement on a
beam splitter may be advantageously used for generating new classes of
quantum states\cite{a28,a28a}. Dakna's group used conditional measurement on
the BS to generate cat-like state\cite{a29}. Podoshvedov et al\cite{a27}
proposed optical scheme for generating both a displaced photon and a
displaced qubit via conditional measurement. In Ref.\cite{a30}, they
proposed to create arbitrary Fock states via conditional measurement on the
BS. In addition, conditional output measurement on the BS may be used to
produce photon-added states for a large class of signal-mode quantum states,
such as thermal state, coherent state, and squeezed states\cite{a31}.
Similarly, photon-subtracted states can be produced by means of conditional
measurement on the BS\cite{a32}. Therefore, based on conditional measurement
on the BS, it is possible to generate and manipulate various nonclassical
optical fields in a real laboratory.

In this paper, we\ study the Hermite polynomial excited squeezed vacuum
state (HESVS), a kind non-Gaussian quantum state, generated by conditional
output measurement on a BS. The calculations show that when a two-mode
squeezed vacuum state (TMSVS) is injected in the input channels and the
photon number of the mode in one of the output channels is measured, then
the mode in the other output channel is prepared in a conditional state that
has the typical features of a Hermite polynomial excited squeezed state. To
exhibit the nonclassical properties of this conditional state, we mainly
analyze the states in terms of the photon number distribution,
sub-Poissonian distribution, quadrature component distribution, and
Quasi-probability distribution including the Wigner function(WF) and Husimi
function(HF). The paper is organized as follows. Section 2 presents the
basic scheme for generation of the HESVS and its normalization related to
Legendre polynomial. The nonclassical properties of the HESVS are
analytically and numerically studied in Section 3-4. The results indicate
that the conditional HESVS is strongly noncassical and non-Gaussian due to
the presence of the partial negative WF. Finally, a summary and concluding
remarks are given in Section 5.

\section{Generation of Hermite polynomial excited squeezed state}

It is well known that the input-output relations at a lossless beam splitter
can be characterized by the SU(2) Lie algebra. In the Schr\"{o}dinger
picture, the role played by the beam splitter (BS) upon the input state $%
\rho _{in}$\ results in the output state
\begin{equation}
\rho _{out}=\hat{B}\rho _{in}\hat{B}^{\dag },  \label{1.1}
\end{equation}%
where $\hat{B}=\exp \left[ \theta \left( a^{\dag }b-ab^{\dag }\right) \right]
$\ corresponds to the unitary operator in terms of the creation
(annihilation) operator $a^{\dag }$($a$) and $b^{\dag }$($b$)\ for mode $a$
and $b$, whose transformations satisfy\cite{a30}
\begin{align}
\hat{B}a\hat{B}^{\dag }& =a\cos \theta -b\sin \theta ,  \notag \\
\hat{B}b\hat{B}^{\dag }& =a\sin \theta +b\cos \theta .  \label{1.1a}
\end{align}%
Moreover, $\cos \theta $ and $\sin \theta $\ are\ the transmittance and
reflectance\ of the beam splitter, respectively. Note that the globe phase
factor of BS may be omitted without loss of generality. For the sake of
simplicity, we also assume that $\theta $ is tunable in the range of $\left[
0,\pi /2\right] $. Under special circumstances, when $\theta =0$ or $\theta
=\pi /2$, the BS corresponds to the cases of total transmission and total
reflection, respectively. For $\theta =\pi /4$, the BS is just the
symmetrical, i.e. 50/50 BS.

\subsection{Hermite polynomial excited squeezed state}

A two-mode squeezed vacuum state (TMSVS) is the correlated state of
two field modes $a$ and $b$ (signal and idle) that can be generated
by a nonlinear medium. Theoretically, the TMSVS is obtained by
applying the
unitary operator $S_{2}\left( r\right) $ on the two-mode vacuum,%
\begin{equation}
\left\vert \Psi \right\rangle _{ab}=S_{2}\left( r\right) \left\vert
0,0\right\rangle =\cosh ^{-1}re^{a^{\dag }b^{\dag }\tanh r}\left\vert
0,0\right\rangle ,  \label{1.2}
\end{equation}%
where $S_{2}\left( r\right) =\exp \left[ r\left( a^{\dag }b^{\dag
}-ab\right) \right] $ is the two-mode squeezed operator and the values of $r$
determines the degree of squeezing. The larger $r$, the more the state is
squeezed. Especially, when $r=0$, $\left\vert \Psi \right\rangle _{ab}$
reduces to two-mode vacuum state $\left\vert 0,0\right\rangle $.
\begin{figure}[tbp]
\label{Fig1-1} \centering\includegraphics[width=0.9\columnwidth]{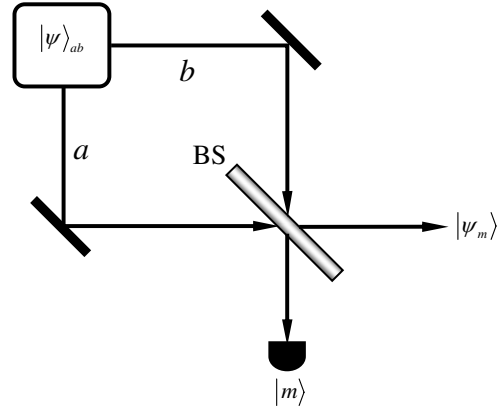}
\caption{{\protect\small Preparation scheme of HPESVS. When a TMSVS is mixed
by a beam splitter and the number of photons }$\left\vert m\right\rangle $%
{\protect\small is measured in one of the output channels, then the
conditional quantum state in the other output channel is generated.}}
\end{figure}

The conceptual scheme of the experimental setup is depicted in Fig.1. The
two input modes prepared in the two-mode squeezed state ($\rho
_{in}=\left\vert \Psi \right\rangle _{ab}\left\langle \Psi \right\vert $) is
mixed at BS, so the output-state density operator can be given by $\rho
_{out}=\hat{B}\left\vert \Psi \right\rangle _{ab}\left\langle \Psi
\right\vert \hat{B}^{\dag }.$ In fact the output modes in $\rho _{out}$ are
generally highly correlated. When the photon number of the mode in the
second output channel is measured and $m$ photons are detected, then the
mode in the first output channel is prepared in a quantum state, whose
density operator $\rho _{out}^{a}$ reads as%
\begin{equation}
\rho _{out}^{a}=N_{m}^{-1}\left. _{b}\left\langle m\right\vert \hat{B}%
\left\vert \Psi \right\rangle _{ab}\left\langle \Psi \right\vert \hat{B}%
^{\dag }\left\vert m\right\rangle _{b}\right. =\left\vert \Psi
_{m}\right\rangle \left\langle \Psi _{m}\right\vert ,  \label{1.3}
\end{equation}%
where $\left\vert \Psi _{m}\right\rangle $\ is the normalized output
conditional state (a pure state) and $N_{m}$\ is the normalization factor
determined by $\mathrm{Tr}\left( \rho _{out}^{a}\right) =1$.

Next, using the integration of the TMSVS\cite{a33},
\begin{equation}
\left\vert \Psi \right\rangle _{ab}=\frac{1}{\sinh r}\int \frac{d^{2}\alpha
}{\pi }e^{-\left\vert \alpha \right\vert ^{2}/\tanh r+\alpha a^{\dag
}+\alpha ^{\ast }b^{\dag }}\left\vert 0_{a},0_{b}\right\rangle ,
\label{1.3a}
\end{equation}%
where $\left\vert 0_{a},0_{b}\right\rangle =\left\vert 0\right\rangle
_{a}\otimes \left\vert 0\right\rangle _{b}$ is two-mode vacuum state, and
the transformation relation in Eq.(\ref{1.1a}), after some algebra we derive
that%
\begin{align}
& _{b}\left\langle m\right\vert \hat{B}\left\vert \Psi \right\rangle _{ab}
\notag \\
& =\frac{1}{\sinh r}\int \frac{d^{2}\alpha }{\pi }e^{-\left\vert \alpha
\right\vert ^{2}/\tanh r}e^{\left( \alpha \cos \theta +\alpha ^{\ast }\sin
\theta \right) a^{\dag }}\left\vert 0\right\rangle _{a}  \notag \\
& \times \frac{1}{\sqrt{m!}}\frac{\partial ^{m}}{\partial \tau ^{m}}e^{\tau
\left( \alpha ^{\ast }\cos \theta -\alpha \sin \theta \right) }|_{\tau =0}
\notag \\
& =\frac{1}{\cosh r\sqrt{m!}}\frac{\partial ^{m}}{\partial \tau ^{m}}e^{%
\frac{\mu }{2}a^{\dag 2}-\frac{\mu }{2}\tau ^{2}+\tau \nu a^{\dag
}}\left\vert 0\right\rangle _{a}|_{\tau =0},  \label{1.3b}
\end{align}%
where $\mu =\sin 2\theta \tanh r$ and $\nu =\cos 2\theta \tanh r$. Hence the
output conditional state $\left\vert \Psi _{m}\right\rangle $ is explicitly
expressed as
\begin{equation}
\left\vert \Psi _{m}\right\rangle =\Omega _{m}^{1/2}H_{m}\left( \frac{\nu
a^{\dag }}{\sqrt{2\mu }}\right) S_{1}\left( \lambda \right) \left\vert
0\right\rangle   \label{1.4}
\end{equation}%
with $\Omega _{m}=\mu ^{m}\cosh \lambda /(2^{m}m!N_{m}\cosh ^{2}r)$, where
we have used the generating function of the single-variable $m$-order
Hermite polynomial $H_{m}\left( x\right) =\partial _{\tau }^{m}e^{2x\tau
-\tau ^{2}}|_{\tau =0}\ $and the expression of single-mode squeezed vacuum $%
S_{1}\left( \lambda \right) \left\vert 0\right\rangle =\cosh ^{-1/2}\lambda
e^{\left( \tanh \lambda /2\right) a^{\dag 2}}\left\vert 0\right\rangle $
with the single-mode squeezed operator $S_{1}\left( \lambda \right) =\exp %
\left[ \lambda \left( a^{\dag 2}-a^{2}\right) /2\right] $. Eq.(\ref{1.4})
indicates that the conditional state $\left\vert \Psi _{m}\right\rangle $ is
actually a single-mode $m$-order Hermite polynomial excited squeezed vacuum
state. It is worth noticing that the degree of squeezing $\lambda $ of the
conditional state is not the same squeezing parameter $r$ of the TMSVS but
related to the parameters of the TMSVS and the BS satisfying $\tanh \lambda
=\sin 2\theta \tanh r$. For the symmetrical case, i.e., $\theta =\pi /4$, $%
\hat{B}\left\vert \Psi \right\rangle _{ab}|_{\theta =\pi /4}=S_{1a}\left(
r\right) \left\vert 0\right\rangle \otimes S_{1b}\left( -r\right) \left\vert
0\right\rangle $ is just the product state of two separate single-mode
squeezed vacuum state. In this case, when we detect $m$ photons in the
second output channel, the conditional state is always $S_{1}\left( r\right)
\left\vert 0\right\rangle $ with the same squeezed parameter $r$. This is
that the effect of the symmetrical BS splits the entangled TMSVS into two
independent single-mode squeezed vacuum state. Note that when no photons are
detected, $m=0$, then $\left\vert \Psi _{m}\right\rangle $ also reduces to $%
S_{1}\left( \lambda \right) \left\vert 0\right\rangle $.

\subsection{Normalization via probability of such event}

In addition, the normalization factor $N_{m}$ is determined by the
probability $p\left( m\right) $ of such an event given by%
\begin{eqnarray}
N_{m} &=&p\left( m\right)  \notag \\
&=&\mathrm{Tr}\left( \left. _{b}\left\langle m\right\vert \hat{B}\left\vert
\Psi \right\rangle _{ab}\left\langle \Psi \right\vert \hat{B}^{\dag
}\left\vert m\right\rangle _{b}\right. \right)  \notag \\
&=&\frac{1}{m!\sqrt{A}\cosh ^{2}r}\allowbreak \frac{\partial ^{2m}}{\partial
s^{m}\partial \tau ^{m}}e^{-B_{1}s^{2}/2-B_{1}\tau ^{2}/2+\allowbreak
B_{2}s\tau }\allowbreak |_{s=\tau =0}  \notag \\
&=&\frac{\left( -\sqrt{B_{3}}\right) ^{m}}{\cosh ^{2}r\sqrt{A}}P_{m}\left(
\sqrt{B_{4}}\right)  \label{1.9}
\end{eqnarray}
where we have set $A=1-\mu ^{2}$, $B_{1}=\mu /\left( A\cosh ^{2}r\right) $, $%
B_{2}=\nu ^{2}/A$, $B_{3}=\left( \tanh ^{4}r-\mu ^{2}\right) /A$, $B_{4}=\nu
^{4}/\left( A^{2}B_{3}\right) $, and in the last step we have used the
formula of $m$-order Legendre polynomial P$_{m}\left( x\right) $, i.e.,%
\begin{equation}
\frac{\partial ^{2m}}{\partial t^{m}\partial \tau ^{m}}e^{-t^{2}-\tau ^{2}+%
\frac{2x}{\sqrt{x^{2}-1}}\tau t}|_{t,\tau =0}=\frac{2^{m}m!}{\left(
x^{2}-1\right) ^{m/2}}\text{P}_{m}\left( x\right) .  \label{1.10}
\end{equation}%
Especially, when $r=0$, $\left\vert \Psi \right\rangle \rightarrow
\left\vert 0,0\right\rangle $, there is no photons in the output channels.
In this case, there is no necessary to making conditional measurement. So
the event is happen only for $r\neq 0$. When $\theta =0$ or $\theta =\pi /2$%
, leading to $\mu =0$ and $\nu =\tanh r$ or $\nu =-\tanh r$ then $A=1$, $%
B_{1}=0$, $B_{2}=\tanh ^{2}r$, $B_{3}=\tanh ^{4}r$, and $B_{4}=1$, $%
N_{m}|_{\theta =0or\theta =\pi /2}=\tanh ^{2m}r/\cosh ^{2}r$ and $\left\vert
\Psi _{m}\right\rangle $ is just the Fock $\left\vert m\right\rangle $,
which is rational because of the inherent properties of the TMSVS. If $%
\theta =\pi /4$, the BS is just the symmetrical, i.e., 50/50 BS, leading to $%
\mu =\tanh r,$ $\nu =0$ then $A=1-\tanh ^{2}r$, $B_{1}=-\frac{\tanh r}{2}$, $%
B_{2}=0$, then the output states is the the product of two independent
single-mode SVS.

According to Eq.(\ref{1.9}), we discuss the probability of observing such a
conditional $m$-order HPESVS. In Fig.2 the probability $p\left( m\right) $\
is plotted for two parameter values of the BS. For a given transmittance of
BS, $p\left( m\right) $ as a function of the input squeezing parameter $r$
can attain a maximum and the maximum is shifted towards larger values of $r$
when $m$ is increased (see Fig.2a). In this figure, for each $r$, we should
use the transmittance of the BS in a way to optimize the success
probability. The ideal procedure would use the appropriate transmittance for
each value of $r$ and $m$. By tuning the parameters of the interaction,
namely, the control parameter of the BS, the squeezed parameter $r$ of the
TMSVS, and the photon number of conditional measurement $m$, the HPESVS may
be modulated, generating a wide range of nonclassical phenomena, as
described below.
\begin{figure}[tbp]
\label{Fig2-1} \centering\includegraphics[width=0.9\columnwidth]{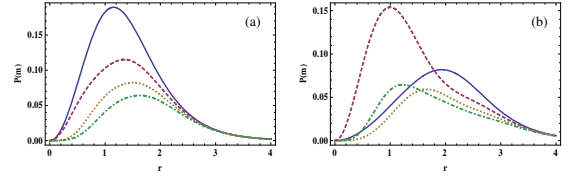}
\caption{{\protect\small The probability of producing }$\left\vert \Psi
_{m}\right\rangle $ {\protect\small is shown as a function of the parameter }%
$r$ {\protect\small of the TMSVS for two parameter values of the BS [(a) }$%
\protect\theta =\protect\pi /7${\protect\small ; (b) }$\protect\theta =2%
\protect\pi /7${\protect\small ] and various values of }$m${\protect\small ,
where }$m=1,2,3,4${\protect\small correspond to the solid, dashed, dotted
and dotdashed lines, respectively.}}
\end{figure}

\section{Observable nonclassical effects of the conditional HPESVS}

To study the nonclassical properties of the conditional states in more
detail, we shall calculate the photon number distribution, sub-Poissonian
distribution, and quadrature component distribution.

\subsection{Photon number distribution}

The photon number distribution (PND), the probability of finding $n$
photons, is a key characteristic of every quantum state. Recalling Eq.(\ref%
{1.4}), the PND of the conditional HPESVS reads as
\begin{equation}
P\left( n|m\right) =\left\vert \left\langle n\right. \left\vert \Psi
_{m}\right\rangle \right\vert ^{2}.  \label{2.1}
\end{equation}%
Using the unnormalized coherent $\left\vert z\right\rangle =\exp [za^{\dag
}]\left\vert 0\right\rangle $, leading to $\left\vert n\right\rangle =\frac{1%
}{\sqrt{n!}}\frac{\partial ^{n}}{\partial z^{n}}\left\vert z\right\rangle
|_{z=0}$, and combining with Eq.(\ref{1.4}), we finally obtain
\begin{align}
& P\left( n|m\right)   \notag \\
& =\frac{1}{m!n!N_{m}\cosh ^{2}r}\allowbreak \left\vert \frac{\partial ^{2m}%
}{\partial \tau ^{m}\partial s^{n}}e^{\frac{\mu }{2}s^{2}-\frac{\mu }{2}\tau
^{2}+\nu s\tau }|_{s=\tau =0}\right\vert ^{2}  \notag \\
& =\frac{m!n!}{N_{m}\cosh ^{2}r}\allowbreak \left\vert \sum_{g=0}^{\min
[m,n]}\frac{\left( -1\right) ^{\frac{m-g}{2}}\left( \allowbreak \frac{\mu }{2%
}\right) ^{\frac{m+n-2g}{2}}\nu ^{g}}{\left( \frac{n-g}{2}\right) !\left(
\frac{m-g}{2}\right) !g!}\right\vert ^{2},  \label{2.2}
\end{align}%
where we have used $\frac{\partial ^{m}}{\partial x^{m}}x^{n}|_{x=0}=m!%
\delta _{mn}$ ($\delta _{mn}$ is Krocher function) and $N_{m}$ is given in
Eq.(\ref{1.9}). In the summation of Eq.(\ref{2.2}), the value of $g$ must
make $\frac{n-g}{2}$\ and $\frac{m-g}{2}$ be integer. To see clearly the
variation of the PND, in Fig.3 we plot the bar graph of the PND for the
conditional HPESVS with different values of parameters $m$, $\theta $, and $r
$. From Fig.3 we easily see that when the number $m$ of the conditional
measurement is odd (even), then the photon-number distribution is nonzero
only for odd (even)\ photon numbers. The probability $P\left( n|m\right) $\
with different parity between $m$ and $n$ is zero. For given $m$ and $\theta
$, the bigger the squeezing parameter $r$, the wider the distribution (see
Figs.3(a) and 3(b)).
\begin{figure}[tbp]
\label{Fig3-0} \centering \includegraphics[width=0.9\columnwidth]{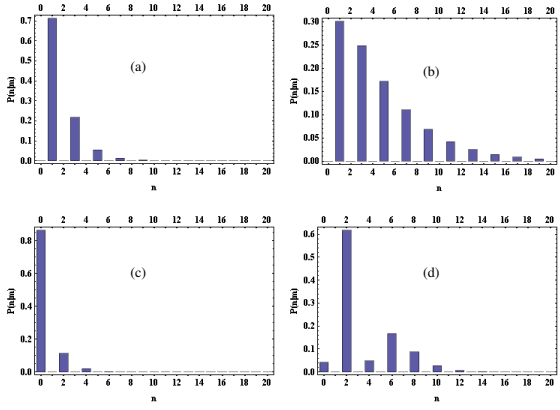}
\caption{{\protect\small Photon-number distribution of the conditional
HPESVS for (a) }$m=1,\protect\theta =2\protect\pi /7,r=0.5${\protect\small ;
(b) }$m=1,\protect\theta =2\protect\pi /7,r=1.0${\protect\small ; (c) }$m=4,%
\protect\theta =2\protect\pi /7,r=1.0${\protect\small ; (d) }$m=4,\protect%
\theta =3\protect\pi /7,r=1.0${\protect\small , respectively.}}
\end{figure}

\subsection{Sub-Poissonian distribution}

In order to study the photon-number statistics of this conditional state, we
first calculate $\left\langle a^{k}a^{\dag l}\right\rangle =\left\langle
\Psi _{m}\right\vert a^{k}a^{\dag l}\left\vert \Psi _{m}\right\rangle $.
Using the completeness of coherent state $\int \frac{d^{2}\alpha }{\pi }%
\left\vert \alpha \right\rangle \left\langle \alpha \right\vert =1$ as well
as $y^{k}=\frac{\partial ^{k}}{\partial t^{k}}e^{ty}|_{t=0}$ yields%
\begin{align}
& \left\langle a^{k}a^{\dag l}\right\rangle   \notag \\
& =\frac{1}{N_{m}m!\cosh ^{2}r\sqrt{A}}\frac{\partial ^{2m}}{\partial
s^{m}\partial \tau ^{m}}e^{-B_{1}s^{2}/2-B_{1}\tau ^{2}/2+B_{2}\tau s}
\notag \\
& \times \frac{\partial ^{k+l}}{\partial x^{k}\partial y^{l}}e^{\left[ \mu
\nu \left( xs+y\tau \right) +\nu \left( ys+x\tau \right) +\mu \left(
x^{2}+y^{2}\right) /2+xy\right] /A}|_{s=\tau =x=y=0}.  \label{2.3}
\end{align}%
Thus, the mean photon number
\begin{equation}
\left\langle n\right\rangle =\left\langle a^{\dag }a\right\rangle
=\left\langle aa^{\dag }\right\rangle -1,  \label{2.4}
\end{equation}%
can be determined by Eq.(\ref{2.3}) with $k=l=1$. Examples are shown in
Figs.4(a) and 4(c). We see that when $\theta =\pi /5$, the number of photons
that can be found in $\left\vert \Psi _{m}\right\rangle $ increases with $m$
for the given larger $r$, and $\left\langle n\right\rangle $ as a function
of $\theta $ is symmetric distribution for $\theta =\pi /4$. This is simply
a consequence of the BS transformation. In particular, when no photons are
detected, $m=0$, then $\left\langle n\right\rangle =\sinh ^{2}r$ reduces to
the mean photon number of single-mode squeezed vacuum state.

A measure of the deviation of the photon number distribution from a
Poissonian is the Mandel $Q$ factor defined by\cite{a34}%
\begin{align}
Q & =\frac{\left \langle n^{2}\right \rangle -\left \langle n\right \rangle
^{2}}{\left \langle n\right \rangle }-1  \notag \\
& =\frac{\left \langle a^{2}a^{\dagger2}\right \rangle -\left \langle
aa^{\dagger}\right \rangle ^{2}-2\left \langle aa^{\dagger}\right \rangle +1%
}{\left \langle aa^{\dagger}\right \rangle -1},  \label{2.5}
\end{align}
It holds that $Q\geqslant0$\ and the equality is achieved for the Fock
state. The light is sub-Poissonian when the photon-number variance $%
\left
\langle n^{2}\right \rangle -\left \langle n\right \rangle ^{2}$ is
less than $\left \langle n\right \rangle $. This is indicated by a negative
value of $Q$. The statistics are Poissonian when $Q=0$, and super- (sub-)
Poissonian if $Q>0$ ($Q<0$). \
\begin{figure}[ptb]
\label{Fig3-1} \centering \includegraphics[width=1.0\columnwidth]{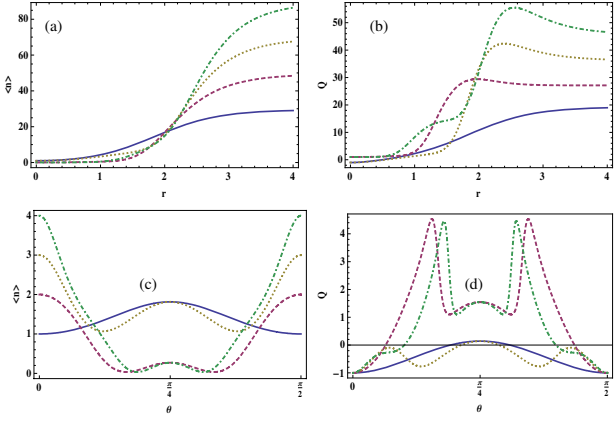}
\caption{{\protect\small (a) Mean photon number }$\left \langle
n\right
\rangle $ {\protect\small and (b) Mandel }${\protect\small Q}$
{\protect\small factor versus }$r$ {\protect\small for different }$m=1$%
{\protect\small (solid line), }$m=2${\protect\small (dashed line), }$m=3$%
{\protect\small (dotted line), and }$m=4${\protect\small (dotdashed line)
with the same }$\protect\theta=\protect\pi /5${\protect\small . (c) Mean
photon number }$\left \langle n\right \rangle $ {\protect\small and (d)
Mandel Q factor versus }$\protect\theta$ {\protect\small for different }$m=1$%
{\protect\small (solid line), }$m=2${\protect\small (dashed line), }$m=3$%
{\protect\small (dotted line), and }$m=4${\protect\small (dotdashed line)
with the same }$r=0.5${\protect\small .}}
\end{figure}
\begin{figure}[ptb]
\label{Fig4-1} \centering \includegraphics[width=1.0\columnwidth]{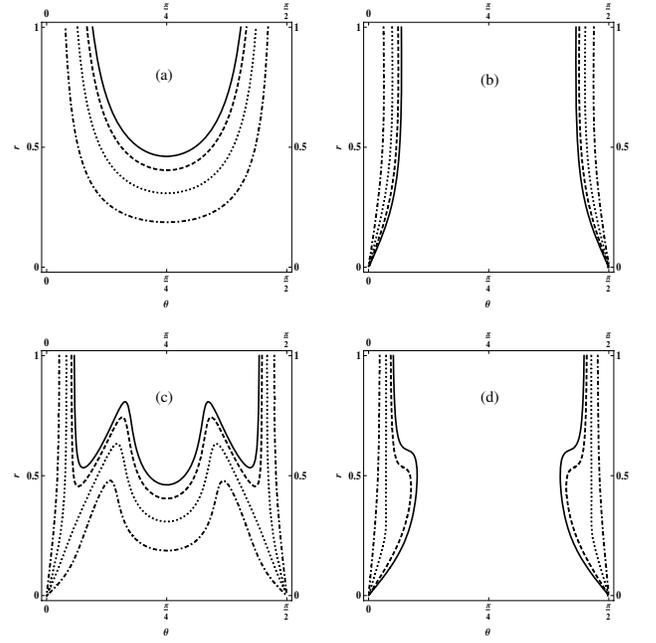}
\caption{{\protect\small The sub-Poissonian properties of the conditional
state with }$Q=0${\protect\small (solid line), }$Q=-0.2${\protect\small %
(dashed line), }$Q=-0.5${\protect\small (dotted line), and }$Q=-0.8$%
{\protect\small (dotdashed line) in the plane space of two parameters (}$%
\protect\theta${\protect\small and }$r${\protect\small ) with different
number of conditional measurement: (a) }$m=1${\protect\small ; (b) }$m=2$%
{\protect\small ; (c) }$m=3${\protect\small ; (d) }$m=4${\protect\small ,
respectively.}}
\end{figure}
\

According to Eqs.(\ref{2.3}) and (\ref{2.5}), we plot the variation of $Q$
for HPESVS versus $r$ or $\theta $ for different $m=1,2,3,4$ in Fig.4(b) and
4(d). It is clearly seen that $Q$ as a function of $\theta $ is also
symmetric distribution for $\theta =\pi /4$ and the HPESVS has
sub-Poissonian statistics behavior due to the emergence of the negativity of
$Q$. With the increasing values of $m$, the increasing intrend of $Q$ is
accelerated for larger $r.$ To further exhibit the high nonclassicality,
Fig.5 shows the dependence of this conditional state on $\theta $\ and $r$
for four different $Q$ factors. Especially, we consider first the boundary
case of the Poissonian distribution, $Q=0$ (see the solid line in Fig.5).

\subsection{Quadrature component distribution}

Next we pay attention to the conditional quadrature component distribution
(QCD)\cite{a28a}%
\begin{equation}
P\left( x,\varphi |m\right) =\left\vert \left\langle x,\varphi \right.
\left\vert \Psi _{m}\right\rangle \right\vert ^{2},  \label{3.1}
\end{equation}%
which can be measured in balanced homodyne detection. Here $\left\vert
x,\varphi \right\rangle $ is the eigenstate of the quadrature component $%
X\left( \varphi \right) =\left( ae^{-i\varphi }+a^{\dag }e^{i\varphi
}\right) $, expressed as in the Fock basis%
\begin{equation}
\left\vert x,\varphi \right\rangle =\pi ^{-1/4}e^{-\frac{x^{2}}{2}+\sqrt{2}%
xa^{\dag }e^{i\varphi }-\frac{a^{\dag 2}e^{2i\varphi }}{2}}\left\vert
0\right\rangle .  \label{3.2}
\end{equation}%
Using Eqs.(\ref{1.4}) and (\ref{3.2}) and inserting the completeness of
coherent state $\int \frac{d^{2}\alpha }{\pi }\left\vert \alpha
\right\rangle \left\langle \alpha \right\vert =1$, after integration$%
\allowbreak $, the wave function $\left\langle x,\varphi \right. \left\vert
\Psi _{m}\right\rangle $ reads
\begin{equation}
\left\langle x,\varphi \right. \left\vert \Psi _{m}\right\rangle =\frac{\pi
^{-1/4}\left( \sqrt{\Gamma /2}\right) ^{m}e^{-\frac{\Pi }{2}x^{2}}}{\sqrt{%
N_{m}m!\left( 1+\mu e^{-2i\varphi }\right) }\cosh r}H_{m}\left( \frac{\Delta
}{\sqrt{\Gamma }}x\right) ,  \label{3.3}
\end{equation}%
where we have set $\Theta =1+\mu e^{-2i\varphi }$, $\Pi =\left( 1-\mu
e^{-2i\varphi }\right) /\Theta $, $\Gamma =\left( \mu +e^{-2i\varphi }\tanh
^{2}r\right) /\Theta $, and $\Delta =e^{-i\varphi }\nu /\Theta $.

As a result of Eq.(\ref{3.3}), we easily obtain the conditional QCD defined
by Eq.(\ref{3.1}) and plot the variation of $P\left( x,\varphi |m\right) $
for the HPESVS as a function of $x$ or $\varphi $ for different $m=1,2,3,4$
in Fig.6. Ones see that for $\varphi $ near $\pi /2$ the QCD $P\left(
x,\varphi |m\right) $ with $m>0$ exhibits two separated peaks, wheresas for $%
\varphi $ close to $0$ and $\pi $ an interference pattern is observed.
\begin{figure}[tbp]
\label{Fig6-1} \centering \includegraphics[width=1.0\columnwidth]{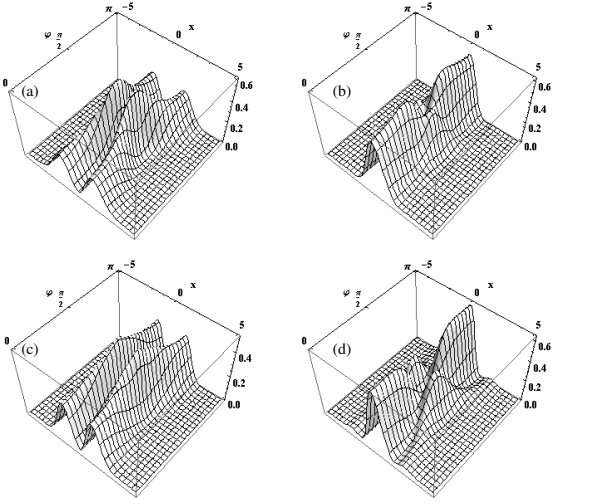}
\caption{{\protect\small Quadrature-component distribution }$P(x,\protect%
\varphi |m)${\protect\small of the conditional state }$\left\vert \Psi
_{m}\right\rangle ${\protect\small for }$\protect\theta =\protect\pi /7,$$%
r=0.5${\protect\small and various numbers }${\protect\small m}$%
{\protect\small of measured photons with (a) }$m=1${\protect\small , (b) }$%
m=2${\protect\small , (c) }$m=3${\protect\small , and (d) }$m=4$%
{\protect\small , respectively.}}
\end{figure}

\section{Quasi-probability distribution of the conditional HPESVS}

Quasi-probability distribution function in the phase space is a very useful
tool for a comprehensive description of the nonclassical state. Thus, in
this section, we shall analytically discuss several quasi-probability
distributions, including Wigner function and Husimi function to characterize
the nonclassicality of the conditional HPESVS.

\subsection{Wigner function}

The WF was first introduced by Wigner in 1932 to calculate quantum
correction to a classical distribution function of a quantum-mechanical
system. The presence of negativity of the WF is a signature of its
nonclassicality\cite{a36,a37}. For a single-mode density operator $\rho$,
the WF in the coherent state representation $\left \vert z\right \rangle $
can be expressed as

\begin{equation}
W(\alpha )=\frac{2e^{2\left\vert \alpha \right\vert ^{2}}}{\pi }\int \frac{%
d^{2}z}{\pi }\left\langle -z\right\vert \rho \left\vert z\right\rangle
e^{-2\left( z\alpha ^{\ast }-z^{\ast }\alpha \right) },  \label{4.1}
\end{equation}%
where $\alpha =\left( x+ip\right) /\sqrt{2}$. The Wigner function of the
conditional state $\rho _{out}^{a}=\left\vert \Psi _{m}\right\rangle
\left\langle \Psi _{m}\right\vert $, can be calculated in a straightforward
way.%
\begin{align}
& W(x,p|m)  \notag \\
& =\frac{2}{\pi N_{m}m!\cosh ^{2}r\sqrt{A}}e^{-2\Xi \left\vert \alpha
\right\vert ^{2}+\frac{2\mu }{A}\alpha ^{2}+\frac{2\mu }{A}\alpha ^{\ast 2}}
\notag \\
& \times \frac{\partial ^{2m}}{\partial \tau ^{m}\partial s^{m}}%
e^{Rs+R^{\ast }\tau -B_{2}\tau s-\frac{B_{1}}{2}s^{2}-\frac{B_{1}}{2}\tau
^{2}}|_{s=\tau =0}  \notag \\
& =\frac{2m!}{\pi N_{m}\cosh ^{2}r\sqrt{A}}e^{-2\Xi \left\vert \alpha
\right\vert ^{2}+\frac{2\mu }{A}\alpha ^{2}+\frac{2\mu }{A}\alpha ^{\ast 2}}
\notag \\
& \times \sum_{l=0}^{m}\frac{\left( -B_{2}\right) ^{l}\left( -B_{1}/2\right)
^{m-l}}{l!\left[ \left( m-l\right) !\right] ^{2}}\left\vert H_{m-l}\left( -%
\frac{R}{\sqrt{2B_{1}}}\right) \right\vert ^{2},  \label{4.2}
\end{align}%
where we have set $\Xi =\left( 1+\mu ^{2}\right) /\left( 1-\mu ^{2}\right) $
and $R=2\nu \left( \alpha -\mu \alpha ^{\ast }\right) /A$. Especially, when
no photons are detected, $m=0$, then $W(x,p|0)$ $\rightarrow \exp \left(
-p^{2}e^{-2\lambda }-x^{2}e^{2\lambda }\right) /\pi $ is a Gaussian form in
phase space, which is just WF of single-mode SVS, as expected.
\begin{figure}[tbp]
\label{Fig7-1} \centering \includegraphics[width=1.0\columnwidth]{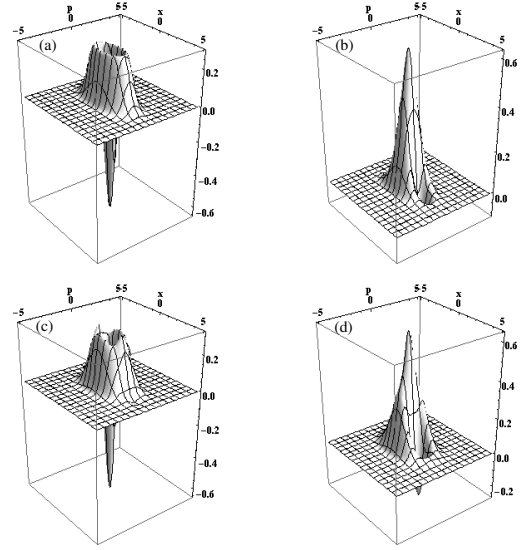}
\caption{{\protect\small Wigner functions }$W(x,p|m)${\protect\small of the
conditional state }$\left\vert \Psi _{m}\right\rangle ${\protect\small for }$%
\protect\theta =\protect\pi /7,$$r=0.5${\protect\small and various numbers m
of measured photons with (a) }$m=1${\protect\small , (b) }$m=2$%
{\protect\small , (c) }$m=3${\protect\small , (d) }$m=4${\protect\small ,
respectively.}}
\end{figure}

The WFs of the conditional HPESVS $\left \vert \Psi_{m}\right
\rangle $ in Fig.7 are plotted for the same parameters as in Fig.6.
The figures indicate that the conditional HPESVS is a noncassical
non-Gaussian state, since the partial negative regions in phase
space are observed in Fig.7. This further demonstrates that
performing the conditional output measurement on a BS is an
effective approach to generate non-Guassian state. In addition, it
is seen from Fig.7 that for odd $m$ there exists a negative valley
in the center region, whereas for even $m$ there exists a main peak.
In fact, for the center region $W(0,0|m)=\frac{2}{\pi}\left(
-1\right) ^{m}$, as expected.

\subsection{Husimi function}

The Husimi function $Q(x,p|m)$ of the state $\left \vert
\Psi_{m}\right
\rangle $ is defined by\cite{a28}
\begin{equation}
Q\left( x,p|m\right) =\frac{1}{\pi}\left \vert \left \langle \beta \right.
\left \vert \Psi_{m}\right \rangle \right \vert ^{2},  \label{4.3}
\end{equation}
where $\left \vert \beta \right \rangle $\ is a coherent state and $%
\beta=\left( x+ip\right) /\sqrt{2}$. Using Eq.(\ref{1.4}), the scalar
product $\left \langle \beta \right. \left \vert \Psi_{m}\right \rangle $
can be easily calculated as follow \
\begin{equation}
\left \langle \beta \right. \left \vert \Psi_{m}\right \rangle =\frac{%
\allowbreak e^{-\frac{\left \vert \beta \right \vert ^{2}}{2}+\allowbreak
\frac{\mu}{2}\beta^{\ast2}}}{\sqrt{m!N_{m}}\cosh r}\frac{\partial^{m}}{%
\partial \tau^{m}}e^{-\tau^{2}\frac{\mu}{2}+\tau \nu \beta^{\ast}}|_{\tau=0},
\label{4.4}
\end{equation}
So we find that $Q\left( x,p|m\right) $\ can be written as%
\begin{equation}
Q\left( x,p|m\right) =\frac{\left( \mu/2\right) ^{m}e^{-\left \vert \beta
\right \vert ^{2}+\allowbreak \frac{\mu}{2}\left( \beta^{2}+\beta^{\ast
2}\right) }}{\pi N_{m}m!\cosh^{2}r}\left \vert \allowbreak H_{m}\left( \frac{%
\nu}{\sqrt{2\mu}}\beta^{\ast}\right) \right \vert ^{2}.  \label{4.5}
\end{equation}
\begin{figure}[ptb]
\label{Fig8-1} \centering \includegraphics[width=1.0\columnwidth]{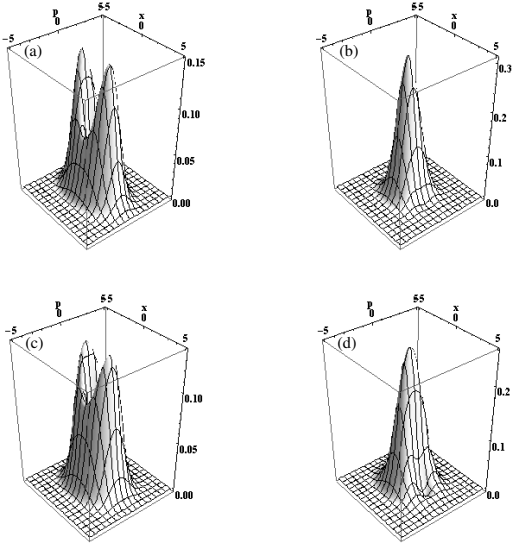}
\caption{{\protect\small Husimi functions }${\protect\small Q(x,p|m)}$
{\protect\small of the conditional state }$\left \vert
\Psi_{m}\right
\rangle $ {\protect\small for }$\protect\theta%
{\protect\small =}\protect\pi{\protect\small /7}${\protect\small , }$%
{\protect\small r=0.5}${\protect\small and various numbers }${\protect\small %
m}$ {\protect\small of measured photons with (a)}${\protect\small m=1}$%
{\protect\small , (b) }${\protect\small m=2}${\protect\small , (c) }$%
{\protect\small m=3}${\protect\small , (d) }${\protect\small m=4}$%
{\protect\small , respectively.} }
\end{figure}

As expected, for $m=0$ the Husimi function is Gaussian, while for
odd $m$ a two-peak structure and for even m a single peak are
observed in Fig.8, whose parameters are the same as WFs in Fig.7.
Note that the Husimi function is a phase-space function that can be
measured in multiport balanced homodyning. Since the Husimi function
can be regarded as a smoothed Wigner function, it is always
non-negative and the oscillating behavior, typical of WF (see
Fig.7), cannot be observed.

\section{Conclusions and Discussions}

In summary, we have shown that Hermite polynomial excited squeezed
states\ can be generated by conditional measurements using a simple
beam splitter scheme. When a two-mode squeezed vacuum state is mixed
by a beam splitter and the number of photons is measured in one of
the output channels, then the conditional quantum state in the other
output channel reveals all properties of a Hermite polynomial
excited squeezed state. Then, we also have numerically analyzed the
conditional HPESVS in terms of the photon-number statistics,
quadrature-component distribution and quasi-probability distribution
such as the Wigner and Husimi functions. The results show that by
tuning the parameters of the interaction, namely, the control
parameter of the BS, the squeezed parameter $r$ of the TMSVS, and
the photon number of conditional measurement $m$, the HPESVS may be
modulated, generating a wide range of nonclassical phenomena. This
further demonstrates that performing the conditional measurement on
a BS is an effective approach to generate non-Guassian state.

\begin{acknowledgments}
This project was supported by the Naitional Nature Science Foundation of
China (Nos.11264018 and 11447002).
\end{acknowledgments}

\end{document}